\documentclass[aps,prc,superscriptaddress,twoside,twocolumn,nofootinbib,showpacs]{revtex4-1}
\usepackage{amsmath,amssymb}
\usepackage{mathtools}
\usepackage{bbold}
\usepackage[usenames, dvipsnames]{xcolor}
\usepackage{braket}
\usepackage{graphicx,bm}
\usepackage{slashed}
\usepackage{booktabs}
\usepackage{tensor}
\usepackage{multirow}
\usepackage{mwe}
\usepackage{changes}
\usepackage{enumerate}
\usepackage{stackrel}
\usepackage{physics}

\allowdisplaybreaks

\usepackage{epsfig}
\usepackage{epstopdf}
\usepackage{subfigure}
\usepackage{float}
\usepackage{soul}

\allowdisplaybreaks

\begin{document}
\title{Generative modeling of nucleon-nucleon interactions}

\author{Pengsheng Wen}
\email{pswen2019@physics.tamu.edu}
\affiliation{Cyclotron Institute, Texas A\&M University, College Station, TX 77843, USA}
\affiliation{Department of Physics and Astronomy, Texas A\&M University, College Station, TX 77843, USA}

\author{Jeremy W. \surname{Holt} }
\email{holt@physics.tamu.edu}
\affiliation{Cyclotron Institute, Texas A\&M University, College Station, TX 77843, USA}
\affiliation{Department of Physics and Astronomy, Texas A\&M University, College Station, TX 77843, USA}

\author{Maggie Li}
\email{mll285@cornell.edu}
\affiliation{
Cornell University, Ithaca, NY 14853, USA}

\begin{abstract}
Developing high-precision models of the nuclear force and propagating the associated uncertainties in quantum many-body calculations of nuclei and nuclear matter remain key challenges for ab initio nuclear theory. In the present work we demonstrate that generative machine learning models can construct novel instances of the nucleon-nucleon interaction when trained on existing potentials from the literature. In particular, we train the generative model on nucleon-nucleon potentials derived at second and third order in chiral effective field theory and at three different choices of the resolution scale. We then show that the model can be used to generate samples of the nucleon-nucleon potential drawn from a continuous distribution in the resolution scale parameter space. The generated potentials are shown to produce high-quality nucleon-nucleon scattering phase shifts. This work provides an important step toward a comprehensive estimation of theoretical uncertainties in nuclear many-body calculations that arise from the arbitrary choice of nuclear interaction and resolution scale. Source code for this project can be found at https://github.com/pswen2019/Glow-nuclear-potential.git.
\end{abstract}

\maketitle
{\it Introduction:}
High-precision models of the nuclear interaction are essential not only for explaining the structure and dynamics of atomic nuclei \cite{hebeler15,Ekstrom:2022yea} but also for describing the properties of hot and dense matter in extreme astrophysical environments, such as neutron stars, core-collapse supernovae, and neutron star mergers \cite{Hebeler:2010jx,Drischler:2021kxf}. 
In the present work, we show that modern machine learning generative models have the ability to learn salient features of the NN interaction, reconstruct distributions of potentials, and create novel and physically reasonable instances of the nuclear potential. Generative models have become a popular research topic in machine learning due to their ability to generate new data based on existing data distributions, which has many practical applications to image and speech synthesis \cite{GAN2014, VAE2022, diffusion2020, GlowTTS2020}. In the present work, we apply generative models as a tool for uncertainty quantification in nuclear many-body calculations, which depend on the resolution scale of the employed nuclear force. The work is related to other recent investigations across diverse fields that have explored generative models in the context of the renormalization group and uncertainty quantification \cite{PhysRevResearch.2.023369,ABDAR2021243,yang2024,bohm19}.

Since the nuclear potential arises from quantum chromodynamics (QCD) as the low-energy effective interaction among composite nucleons, there is an inherent uncertainty due to the choice of resolution scale at which nuclear dynamics is resolved \cite{Bogner:2003wn,Bogner:2009bt}. This resolution scale is typically parameterized in the nuclear potential through a momentum-space regulating function parameter $\Lambda$ that demarcates the separation between low-energy and high-energy (unresolved) physics. In principle, physical observables should be independent of the resolution scale, but in practice, the results of nuclear many-body calculations exhibit a moderate residual uncertainty due to this choice \cite{Lynn:2014zia,Holt:2016pjb,Gasparyan:2022isg}. In recent years, chiral effective field theory (EFT) \cite{weinberg79,epelbaum09rmp,machleidt_chiral_2011} has emerged as a suitable tool to systematically study not only the scale dependence of nuclear interactions but also to estimate uncertainties due to missing physics through an analysis of effective field theory truncation errors \cite{Epelbaum:2014efa,furnstahl15_jpg,furnstahl15prc,wesolowski_bayesian_2016,carlsson_uncertainty_2016,Drischler:2017wtt,Wesolowski:2018lzj,drischler20,Drischler:2020yad,Drischler:2021kxf, PhysRevLett.125.202702, PhysRevC.96.024003}. In the framework of chiral EFT, a specific nuclear interaction is obtained by first defining the high-momentum regulating function and then fitting to experimental scattering and bound state data the low-energy constants (LECs) of the theory that characterize unresolved short-distance physics \cite{entem_peripheral_2015,carlsson_uncertainty_2016}.

Nuclear potentials from chiral EFT are typically fitted at only a few select values of $\Lambda$, from which one can obtain a qualitative estimate of the resolution scale uncertainties. For statistical inference, however, one requires the ability to draw samples of the nucleon-nucleon (NN) potential from a continuous distribution in $\Lambda$, which when combined with EFT truncation errors and variations in the LECs consistent with data \cite{PhysRevC.107.014001, PhysRevC.109.064003} can provide a comprehensive assessment of theory uncertainties.
For this purpose, we utilize the Generative Flow (Glow) model \cite{kingma_glow_2018, dinh_density_2017}, which was originally developed in the field of computer vision for generating realistic images and manipulating their attributes. We adapt and refine the Glow model to develop a generative machine learning model for nuclear potentials. Effectively, we will treat the momentum-space matrix elements of the potential in different partial waves as ``images''. We show that the model can recreate the training potentials and generate new physically reasonable nucleon-nucleon potentials over a continuum of cutoff scales. The reliability of the generated potentials is benchmarked by calculating nucleon-nucleon scattering phase shifts. Finally, we show that a combination of the Glow model and Vision Transformer model \cite{dosovitskiy2021image} allows for the extraction of chiral EFT LECs from the generated nuclear potential matrix elements.


{\it Methods:}
The Glow model attempts to learn the properties of a given dataset by constructing an appropriate probability distribution to estimate the probabilities of some features of a sample. Once an appropriate probability distribution is found, the Glow model can generate novel samples by drawing from the distribution through sampling. The Glow model initializes a trainable distribution whose parameters $\bm{\theta}$ are iteratively adjusted such that the likelihood of the distribution can be maximized with respect to $\bm{\theta}$. In practice, one minimizes  \cite{winkler_learning_2019}
\begin{align}
    \mathcal{L}(\bm{\theta})
    &= 
    - \frac{1}{N}\sum_i \log p_{\bm{\theta}}(\bm{x}_i),
\end{align}
where $p_{\bm{\theta}}(\bm{x}_i)$ is the probability of the sample $\bm{x}_i$, $\bm{\theta}$ represents trainable parameters, and $N$ is the number of samples. 

As a flow-based model, Glow couples several layers of transformation functions to form the flow $f$, which transforms a sample $\bm{x}$ from the data space to the latent space variable $\bm{z}$ via
\begin{align}\label{eq:flow}
    \bm{z} 
    &= 
    f_K \circ f_{K-1} \circ \dots  \circ f_{k} \circ \dots \circ f_{1}(\bm{x}),
\end{align}
where $f_k$ is the transformation function at the $k$-th layer. 
The Glow model incorporates multiple-scale architecture \cite{dinh_density_2017} for efficiency purposes. 
A sequence of operations is performed on each sample at every scale. The first operation, squeezing, changes the shape of a sample from $(C, H, W)$ to $(4\times C, H/2, W/2)$, where $C$ is the number of channels, and $H$ and $W$ are the height and width, respectively. At the second operation, $K$ layers of transformation functions are coupled. The sample now is transformed according to Eq.~\eqref{eq:flow}. 
Upon reaching the end of the scale, the sample is again split into two halves along the channel dimension. 
The first half is treated as part of the fully transformed sample $\bm{z}$ in the resulting latent space. 
Its probability will be calculated by a tractable base distribution model whose parameters are the output of neural networks, and contribute to the total probability $q_{\theta}(\bm{z})$. 
The remaining half is employed as the input of the neural networks responsible for constructing the base distribution and will continue progressing to the next scale. 
A tensor with the shape of $(C, H, W)$ traverses through $L$-layer scales 
and finally there are still $(2^{L}\times C, H/2^{L}, W/2^{L})$ elements whose distribution has not been determined. The base distribution to evaluate the probabilities of final elements is built by neural networks whose inputs are the labels of the sample. 
In our work, the base distributions are Gaussian distributions.

To produce samples with new labels, one obtains samples in latent space and transforms them back into data space. For this we apply two different approaches: 1.\ \emph{latent space sampling (LSS)} and 2.\ \emph{latent space interpolation (LSI)}. In LSS we first draw samples $\bm{z}$ in the latent space from a Gaussian which is built according to the labels from the training set. Next, we perform interpolation based on the obtained $\bm{z}$ samples and the new label. Finally, we transform the interpolated sample back into the data space. In LSI the samples of $\bm{z}$ are obtained by transforming given $\bm{x}$s from the data space to the latent space. 
The next two steps are the same as in LSS approach. 
The first approach is suitable for drawing samples from distributions with desired labels, while the second approach can be applied to predict instances of samples based on existing samples.
For nuclear interaction modeling, we view the partial wave matrix elements as the samples for the Glow model. 
Therefore, a nuclear potential can be viewed as a sample with shape $(C, H, W)$, where $C$ is the number of partial wave channels $\alpha$, and $H$ and $W$ are the number of momentum-space mesh points of $p$ and $p'$ for the potential $V_{\alpha}(p, p')$. 

\begin{figure}[t]
    \centering
    \includegraphics[width=\linewidth]{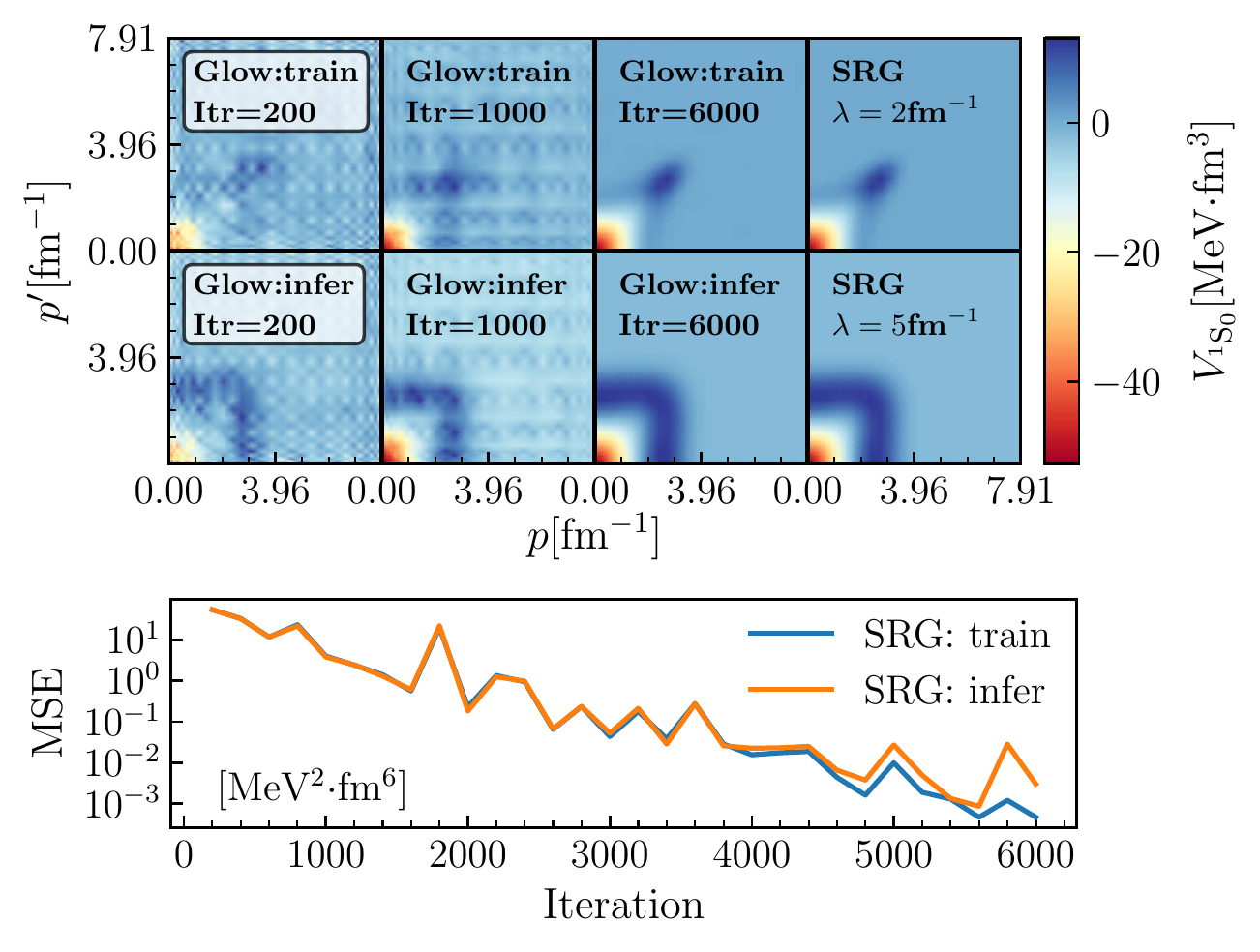}
    \caption{ (Top) $V_{^1{\rm S}_0}$ matrix elements of glow-generated SRG potentials at $\lambda=2$ and $5$ fm$^{-1}$ at different training iterations. 
    The true SRG potential matrix elements are shown on the right side for comparison. 
    (Bottom) The mean squared error, calculated based on the difference of all the elements from all the corresponding potentials.}
    \label{fig:train}
\end{figure}

{\it Results:} We conduct two experiments to investigate the Glow model's capabilities in reconstructing actual nuclear potentials, building distributions, and generating new, realistic ones. Our first experiment employs potentials generated from the Similarity Renormalization Group (SRG) \cite{bogner07,jurgenson_evolution_2009}, where the resolution scale can be freely chosen.
In the SRG, a potential $V_\lambda$ is evolved with a flow parameter $\lambda$ from an initial potential $V$ by a unitary transformation. 
In this paper, the initial potential for the SRG is the next-to-next-to-next-to-leading order (n3lo) chiral nuclear potential of Entem and Machleidt \cite{PhysRevC.68.041001} with momentum-space cutoff $\Lambda=500$\ MeV.
The second experiment attempts to train the Glow model to learn the characteristics of chiral potentials n$\nu$lo$\Lambda$ at different orders $\nu$ in the chiral expansion and different high-momentum cutoff values $\Lambda$.
The SRG potential experiments include three partial-wave channels $C = 3$, and the size of the momentum-space mesh grid is $H = W = 32$. In the chiral potential experiments, we include all partial waves that have associated short-distance contact terms, which requires $C=14$, and we set $H=W=48$ . In addition, $(L, K) = (3, 2)$ for the SRG potential experiments, and $(L, K) = (4, 4)$ for the chiral potential experiments.

To train a Glow model for the SRG, we gather potentials with $\lambda \in \{2, 3, 4, 7, 8, 10, 11, 12\}\, {\rm fm}^{-1}$. 
We then employ the Glow model to deduce potentials with $\lambda = {5, 6, 9, 13}\, {\rm fm}^{-1}$ using LSI throughout the training to evaluate the model's inference capability. When using LSI, the input potentials are the actual SRG potentials used for training. We calculate the mean square error (MSE) by measuring the difference between all the matrix elements of Glow-generated potentials and the actual potentials from SRG. In the bottom panel of Fig.~\ref{fig:train}, we show the MSE as a function of training iteration. During the training process, both the reconstructed potentials (blue line) and the inferred potentials (orange line) converge toward the true potentials generated by the SRG, as indicated by the decreasing MSE. The top panel of Fig.~\ref{fig:train} shows the $\lambda = 2$\,fm$^{-1}$ (training) and $\lambda = 5$\,fm$^{-1}$ (inferred) Glow-generated SRG potential matrix elements in the $^1{\rm S}_0$ partial wave at different iterations (3 columns on the left) and the actual potentials from SRG (rightmost column). 
Initially, the Glow model only generates Gaussian noise. Upon completion of the training, the potentials obtained from the Glow model and the SRG appear indistinguishable. 


Next, we train a separate Glow model based on a set of six chiral potentials with truncation orders $\nu \in \{2, 3\}$ and cutoffs $\Lambda \in \{450, 500, 550\}\, {\rm MeV}$ \cite{PhysRevC.96.024004}. Since the chiral potentials generated by the Glow model are given in terms of their partial-wave matrix elements, we have also trained a Vision Transformer (ViT) model \cite{dosovitskiy2021image, vaswani2017attention} to deduce the associated LECs and value of $\Lambda$ from Glow-generated potentials. 
The ViT model is a widely used machine learning model that is well-suited for regression and classification tasks. In the present case, the ViT model is trained on a large dataset of (unphysical) chiral potentials with LECs randomly sampled from a uniform distribution. 

\begin{figure}[t]
    \centering
    \includegraphics[width=\linewidth]{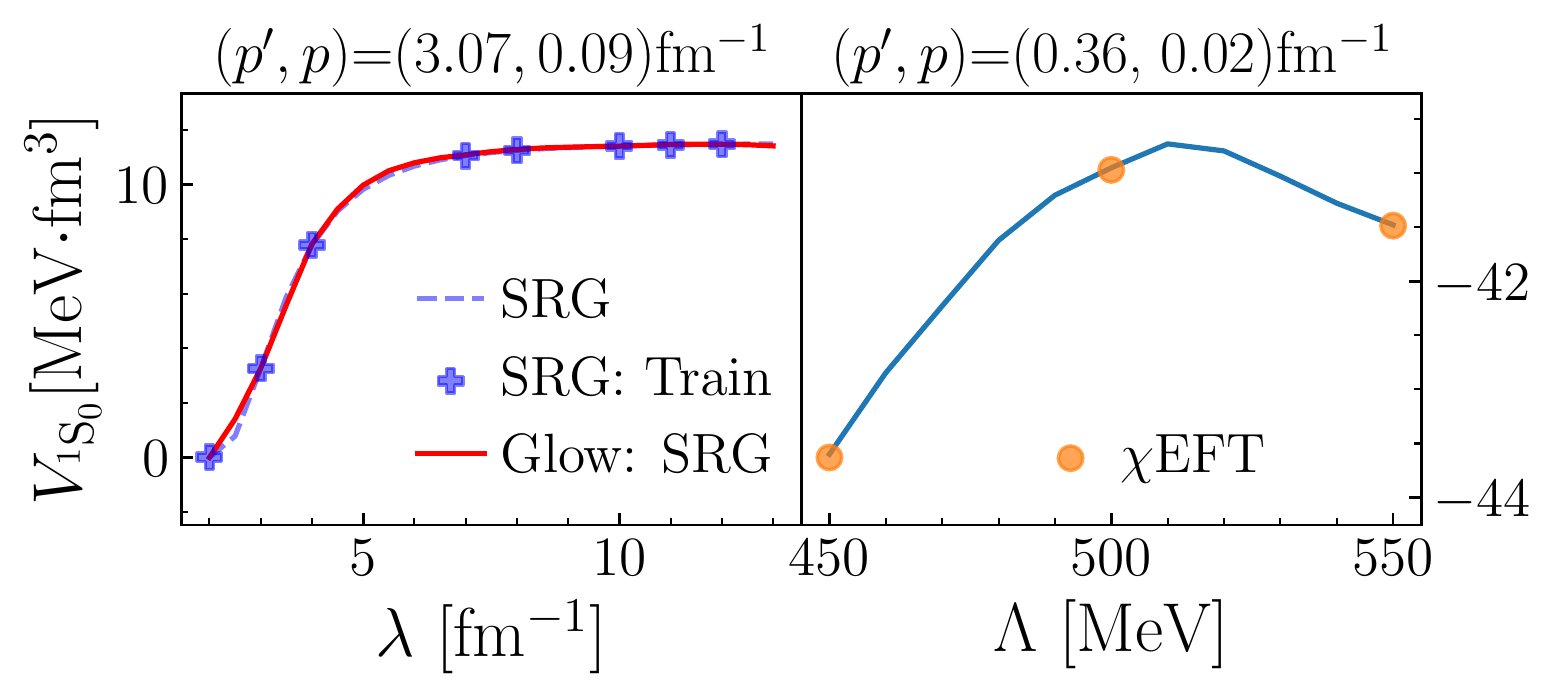}
    \caption{SRG (left) and chiral (right) off-diagonal potential matrix elements in the $^1{\rm S}_0$ partial-wave channel. The plus and dot markers as well as the dashed lines are from the exact potentials, while solid lines are generated by trained Glow models using LSI and LSS for SRG and chiral potentials, respectively.}
    \label{fig:mel}
\end{figure}

\begin{figure*}[t]
    \centering
    \includegraphics[width=0.9\linewidth]{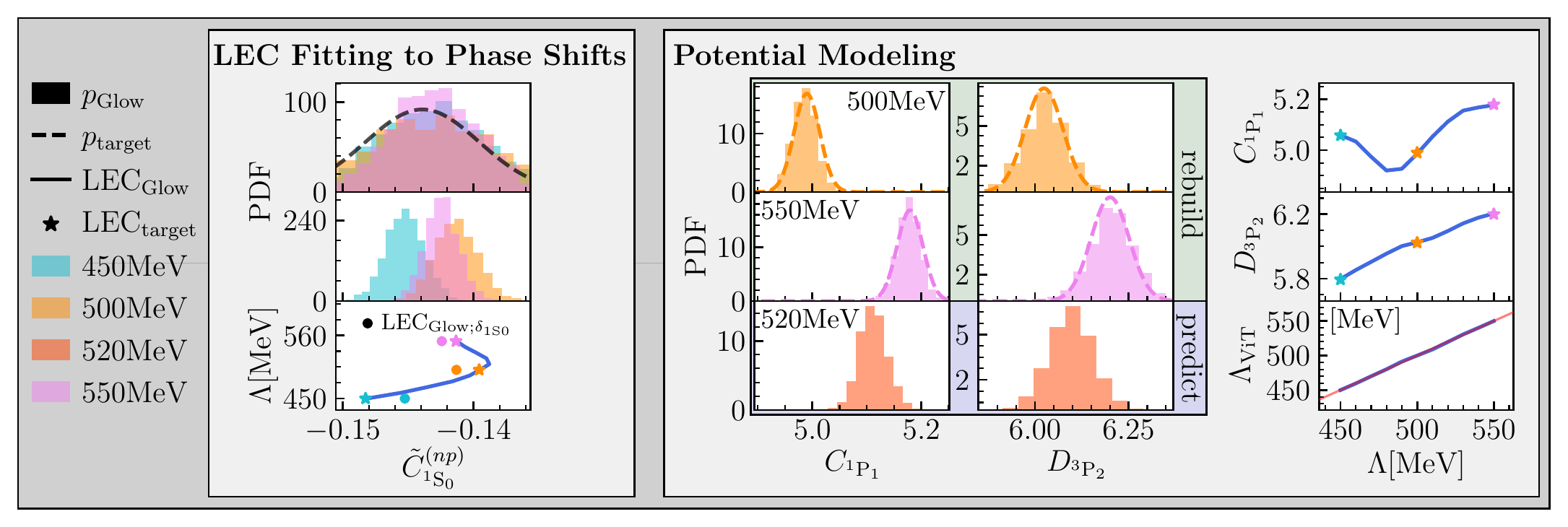}
    \caption{(Left panel) Distributions for $\tilde{C}_{^1{\rm S}_0}$ before (top row) and after (middle row) phase shift training. The mean values of the distributions after phase shift training are shown as circles, while the literature values are shown as stars.
    (Right panel first two columns) LEC distributions predicted using LSS for n3lo chiral potentials with $\Lambda = 500, 550, 520$\,MeV. We show for comparison the input distributions as dashed lines. The rightmost column shows the evolution of the LEC distribution mode as a function of the cutoff, where the stars are the LECs values from the actual chiral n3lo potentials.
    The bottom-right panel shows the ViT predicted cutoff vs.\ the cutoff input label into the Glow model. A red diagonal line (bottom right) is plotted to gauge the quality of the Glow-learned and ViT-extracted cutoffs. }
    %
    \label{fig:lecs}
\end{figure*}

In Fig.~\ref{fig:mel}, we show the cutoff dependence of selected $^1{\rm S}_0$ off-diagonal matrix elements of the Glow-generated SRG potentials (left panel) and chiral potentials (right panel). The Glow-generated SRG potentials are obtained from LSI, where the inputs for LSI are the actual pre-existing potentials, while the Glow-generated Chiral potentials are generated via LSS. In the left panel, the plus symbols mark the actual values of the SRG potential matrix elements at the values of $\lambda$ used for training, which naturally lie exactly on the dashed-blue line that indicates the actual SRG matrix elements as a function of $\lambda$. The matrix elements from the Glow model are shown with the solid red line, which lies nearly on top of the blue dashed line of the actual SRG matrix elements. We note that the Glow model is not naively interpolating between training points, since the points between $\lambda = 4$ and $7$\,fm$^{-1}$ would then be much farther from the blue-dashed SRG line. In addition, from the left panel of Fig.~\ref{fig:mel} we see that the Glow model produces the expected \cite{bogner_low-momentum_2010} decay $V_{\lambda}(p_1, p_2) \approx V(p_1, p_2)e^{-(p_1^2 - p_2^2)^2/\lambda^4}$ of the far off-diagonal matrix element \mbox{$\langle p^\prime = 3.07\,\text{fm}^{-1} | V_{^1{\rm S}_0} | p = 0.09\,\text{fm}^{-1}\rangle$} with respect to $\lambda$. 
Similarly, in the right panel of Fig.\ \ref{fig:mel}, we see that the chiral EFT matrix elements \mbox{$\langle p^\prime = 0.36\,\text{fm}^{-1} | V_{^1{\rm S}_0} | p = 0.02\,\text{fm}^{-1}\rangle$} generated by the Glow model across a continuum of $\Lambda$ values follow a smooth contour that passes through the actual values at $\Lambda = 450, 500, 550$\ MeV.

\begin{figure}[b!]
    \centering
    \includegraphics[width=\linewidth]{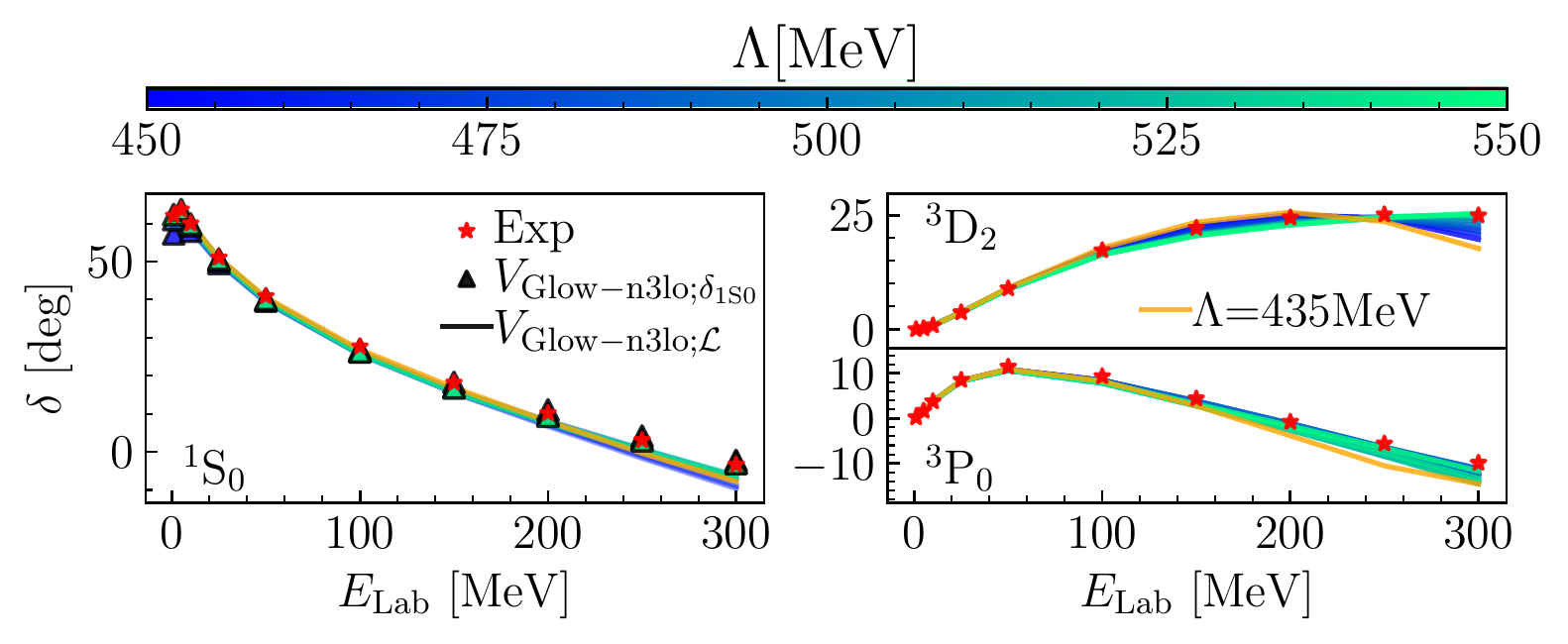}
    \caption{Phase shifts for $np$ scattering. 
    The colored solid lines correspond to Glow-n3lo$\Lambda$ potentials. Red stars correspond to the phase shift analysis \cite{navarro_perez_statistical_2014} of NN scattering data.}
    \label{fig:phase}
\end{figure}

The Glow model can also be used to obtain LEC distributions either (i) starting from scratch by directly fitting to phase shifts or (ii) from existing LEC distributions at different resolution scales. To demonstrate (i), in the left panel of Fig.~\ref{fig:lecs} we start by training a Glow model at three different cutoffs $\Lambda = \{450\, \text{(cyan)}, 500\, \text{(orange)}, 550\,  \text{(magenta)}\}$\,MeV using a wide distribution (top panel, dashed line) for the LEC $\tilde{C}_{^1{\rm S}_0}^{(np)}$. The Glow model is able to rebuild this wide distribution at all three values of $\Lambda$ as shown by the different colored distributions in the top panel. The pre-trained Glow model is then further trained to minimize the difference between the calculated and experimental phase shifts $\delta_{^1{\rm S}_0}$ in the neutron-proton channel up to the laboratory energy of 300 MeV.
As shown in the middle panel of the left subfigure in Fig.\ \ref{fig:lecs}, after the phase shift training, the distributions for the $C_{^1 {\rm S}_0}^{(np)}$ LEC extracted by the ViT model converge separately for the three values of $\Lambda$. Therefore, the Glow model can obtain LEC distributions for different cutoffs from scratch. In the bottom panel of the left subfigure, we show the peak values of $C_{^1 {\rm S}_0}^{(np)}$ determined from the above method (circle symbols) compared to the actual values from the literature (star symbols). The values are not identical, since the literature values were obtained by fitting to experimental scattering data and deuteron properties. Nevertheless, the LEC values are quite similar and show the same trend as $\Lambda$ is varied. Finally, the obtained potentials produce high-quantity phase shifts as shown with the triangle symbols in the left panel of Fig.~\ref{fig:phase}.

In the right panel of Fig.~\ref{fig:lecs} we show the conditional distributions of LECs for given $\Lambda$s in the first two columns, where the potentials are sampled via LSS. Again, we find that the Glow model has the ability to rebuild the LEC distribution at the training values of $\Lambda = 500, 550$\, MeV (top two rows). 
We also show in the third row the predicted distribution $p({\rm LECs}|\Lambda=520\,{\rm MeV})$.
The central point of the predicted distribution is positioned within the range between the centers of $p({\rm LECs}|\Lambda=500\,{\rm MeV})$ and $p({\rm LECs}|\Lambda=550\,{\rm MeV})$. 
Finally, the right column shows the continuous evolution of the peak in the LEC distributions between $\Lambda = 450 - 550$\,MeV, where the actual values are shown as stars for comparison. 
We also show in the bottom-right panel of Fig.~\ref{fig:lecs} the relation between the cutoff $\Lambda_{\rm Glow}$ as the input label for the Glow model and the cutoff $\Lambda_{\rm ViT}$ inferred by the ViT model. The near identity between the two $\Lambda$ values shows that the Glow model can generate chiral potentials with the desired value of $\Lambda$ and that the ViT model can accurately extract $\Lambda$ from the potential matrix elements.

To verify the quality of the chiral potentials generated by the Glow model, in Fig.~\ref{fig:phase} we show the calculated phase shifts in selected partial-wave channels. 
From Fig.~\ref{fig:phase} 
we observe that the Glow-${\rm n}3{\rm lo}\Lambda$ potentials 
give good phase shift results when compared to experimental data used to fit the actual chiral potentials. The ability of the Glow model to extrapolate outside the $\Lambda$ training range is illustrated by the phase shifts for the Glow-generated potential at $\Lambda=435$\,MeV.
Furthermore, even though the values of $\Lambda$ are uniformly sampled over the range $450 < \Lambda < 550$, 
the phase shifts need not be uniformly distributed.

\begin{figure}[t!]
    \centering
    \includegraphics[width=\linewidth]{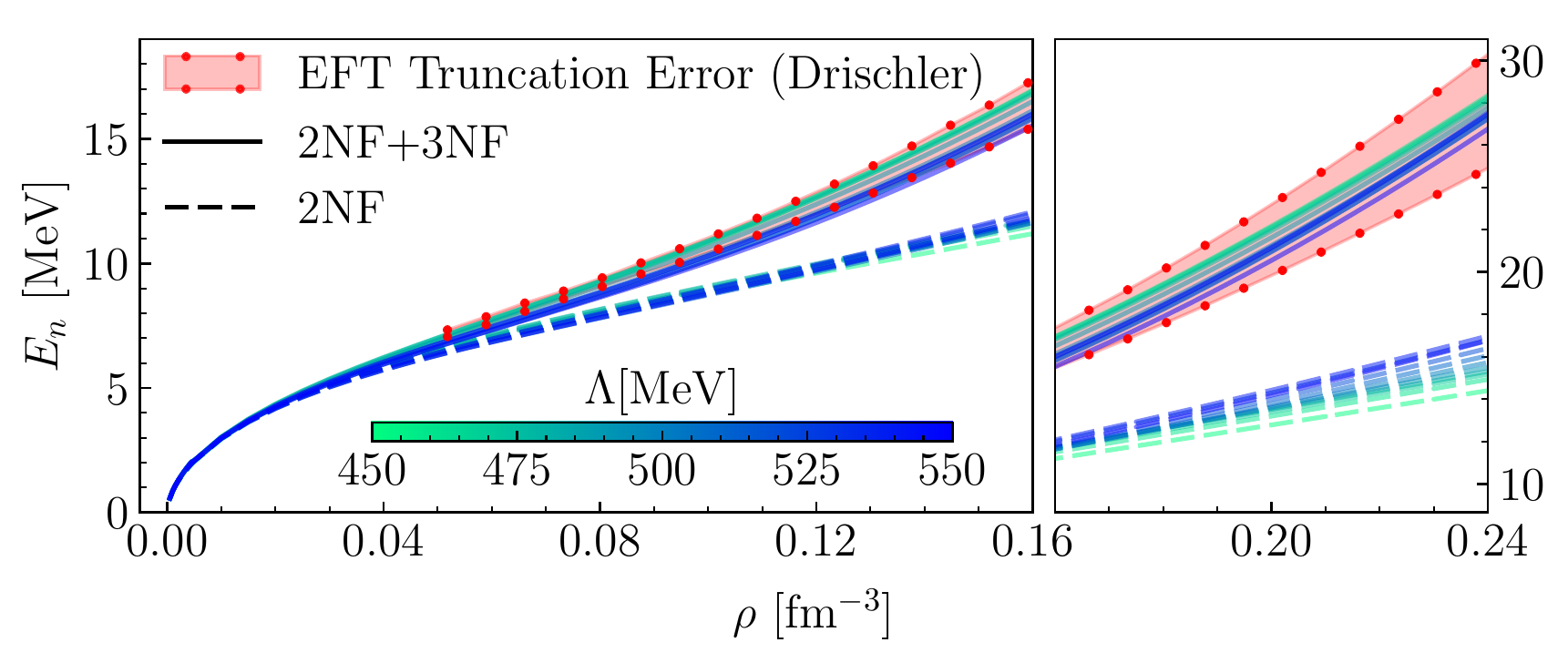}
    \caption{Pure neutron matter equation of state with different cutoffs. The EFT truncation error (red region) from Ref.\ \cite{drischler_quantifying_2020} is plotted for comparison.}
    \label{fig:EOS}
\end{figure}

Finally, as a test case, we study the influence of the resolution scale on the zero-temperature neutron matter equation of state calculated up to second order in perturbation theory, shown in Fig.~\ref{fig:EOS} for two-body forces alone and including the n2lo chiral three-body force. Below saturation density, the uncertainties due to the choice of cutoff scale are larger than those arising from the truncation in the EFT expansion \cite{drischler_quantifying_2020} shown as the red band. However, beyond nuclear saturation density the EFT expansion parameter increases and one observes the EFT truncation errors to grow stronger than those due to the resolution scale. Again, we can find that the distribution of the energy is not a simple uniform distribution with respect to the cutoff. 

{\it Summary and Outlook:}
In this work, we have extended the Glow machine learning model to generate novel instances of the nucleon-nucleon interaction by training the neural network on existing interactions in the literature. We have shown that the Glow model can accurately reconstruct the training nuclear potentials and build a continuous distribution of potentials over a range of resolution scales, all while reproducing nucleon-nucleon scattering phase shift data. The Glow model enables the generation of realistic nucleon-nucleon potentials in a matter of seconds, and therefore it can play an important role for more reliable estimations of nuclear many-body uncertainties that arise due to the arbitrary choice of resolution scale in the nucleon-nucleon interaction. The treatment of nuclear three-body forces within the Glow model is expected to be a straightforward extension of the methods developed in this study and will be pursued in future work. These tools add to the growing body of recent literature exploring machine learning models for uncertainty quantification and the renormalization group.

\begin{acknowledgments}
We thank Takayuki Miyagi for providing us with codes for generating SRG NN potentials. Work supported by the National Science Foundation under Grant Nos.\ PHY1652199 and PHY2209318. Portions of this research were conducted with the advanced computing resources provided by Texas A\&M High Performance Research Computing.
\end{acknowledgments}

%


\end{document}